\documentclass[sigconf]{aamas}

\usepackage{balance} % for balancing columns on the final page
\usepackage{tikz}

\setcopyright{ifaamas}
\acmConference[AAMAS '25]{Proc.\@ of the 24th International Conference
on Autonomous Agents and Multiagent Systems (AAMAS 2025)}{May 19 -- 23, 2025}
{Detroit, Michigan, USA}{A.~El~Fallah~Seghrouchni, Y.~Vorobeychik, S.~Das, A.~Nowe (eds.)}
\copyrightyear{2025}
\acmYear{2025}
\acmDOI{}
\acmPrice{}
\acmISBN{}

\acmSubmissionID{<<173>>}

\title{Shifting Power: Leveraging LLMs to Simulate Human Aversion in ABMs of Bilateral Financial Exchanges, A bond market study}
\author{Alicia Vidler}
\affiliation{
  \institution{UNSW}
  \city{Sydney}
  \country{Australia}
\email{a.vidler@unsw.edu.au}}

\author{Toby Walsh}
\affiliation{
  \institution{UNSW}
  \city{Sydney}
  \country{Australia}
\email{t.walsh@unsw.edu.au}}

\begin{abstract}
    Bilateral markets, such as those for government bonds, involve decentralized and opaque transactions between market makers (MMs) and clients, posing significant challenges for traditional modeling approaches. To address these complexities, we introduce \textbf{TRIBE} an agent-based model augmented with a large language model (LLM) to simulate human-like decision-making in trading environments. TRIBE leverages publicly available data and stylized facts to capture realistic trading dynamics, integrating human biases like risk aversion and ambiguity sensitivity into the decision-making processes of agents. Our research yields three key contributions: first, we demonstrate that integrating LLMs into agent-based models to enhance client agency is feasible and enriches the simulation of agent behaviors in complex markets; second, we find that even slight trade aversion encoded within the LLM leads to a complete cessation of trading activity, highlighting the sensitivity of market dynamics to agents' risk profiles; third, we show that incorporating human-like variability shifts power dynamics towards clients and can disproportionately affect the entire system, often resulting in systemic agent collapse across simulations. These findings underscore the emergent properties that arise when introducing stochastic, human-like decision processes, revealing new system behaviors that enhance the realism and complexity of artificial societies.

\end{abstract}


\newcommand{\BibTeX}{\rm B\kern-.05em{\sc i\kern-.025em b}\kern-.08em\TeX}

\begin{document}

%%% The following commands remove the headers in your paper. For final 
%%% papers, these will be inserted during the pagination process.

\pagestyle{fancy}
\fancyhead{}

%%% The next command prints the information defined in the preamble.
\maketitle 
%%%%%%%%%%%%%%%%%%%%%%%%%%%%%%%%%%%%%%%%%%%%%%%%%%%%%%%%%%%%%%%%%%%%%%%%

\section{Introduction}

\textbf{"Usually there is only one way to be fully rational, but there are many ways to be less rational"} \cite{Holland}

Agent-based models (ABMs) are versatile applications for modeling complex and dynamic systems, particularly suited for bilateral markets like government bond markets. These markets, characterized by direct transactions between two parties without centralized exchanges, present modeling challenges due to their decentralized nature, complex interactions between heterogeneous agents, and lack of transparency.

We introduce the \textit{TRIBE} model, a generative-ABM focused on \textbf{T}rading \textbf{R}elationships, \textbf{I}nteractions, and \textbf{B}ilateral \textbf{E}xchange of assets. \textit{TRIBE} incorporates client agency with dynamically assigned asset distributions and probabilistic trading availability, extending this approach by integrating a large language model (LLM) for more human-like decision-making and negotiation behavior.

Our research yields three key findings. First, we demonstrate the feasibility of using generative ABMs (GABMs) with enhanced client agency, building on work by \cite{park2023generative}, which demonstrates that generative agents can act as human simulacra, and thus advancing the capabilities of financial market simulations. Second, we discover that any LLM prompt suggestion of trade aversion results in a complete cessation of trading, highlighting the sensitivity of market dynamics to agent risk profiles. Third, and most significantly, by incorporating synthetic human client variability via an LLM, we reveal an emergent property where power dynamics shift towards clients, often resulting in significant systemic market collapse across simulations.

We develop a versatile, agent-based simulation of the over-the-counter (OTC) bond market, using LLMs to enrich ABMs with more nuanced, human-like behaviors. Our model focuses on the Australian government bond market, an ideal subject due to its decentralized structure and emphasis on liquidity and flow rather than pricing, building on earlier work by \cite{vidler2024decodingotcgovernmentbond}. We have no reason to suggest that dynamics and results are not generalizable to other similar markets.

We demonstrate the feasibility and benefits of integrating LLMs into GABMs, we advance the field beyond traditional deterministic or simple stochastic approaches used in prior work such as \cite{Cont2023}. Our findings provide valuable insights into the sensitivity of opaque, decentralized markets to agent heterogeneity and decision-making processes, with implications for market design and regulatory policy.

\section{Literature and Recent work}

\subsection{Agent-based models}
The application of agent-based modeling (ABM) to financial markets has evolved significantly, focusing on simulating complex behaviors among market participants. Several prior works have explored ABM applications in trading environments with varying levels of transparency. For instance \cite{vidler2024decodingotcgovernmentbond} apply ABMs to model opaque bilateral bond markets, while \cite{Cont2023} investigates liquidity dynamics in financial market order books.  Recent work by \cite{SamarthNetwork} introduces the concept of \textit{network agency} and the fact that features beyond an agent's direct experience can influence their behaviors.  However, many of these models rely on homogeneous agent classes and observable environments, which are less relevant in over-the-counter (OTC) bond markets where data availability is limited \cite{vadori2022multiagent}. Despite no consensus method for calibrating ABMS \cite{avegliano2019using}, we propose an inductive approach using market structure data such as \cite{vidler2024decodingotcgovernmentbond} and \cite{Pinter2023} to specify agent calibrations. The utility functions and policies utilized within our simulation approach stem from finance literature and regulatory requirements.

\textbf{The Australian Case:}
The Australian government bond market, like other global bond markets, involves a diverse range of participants \cite{RBA}. In 2023, the Australian government debt (bonds issued) exceeded \$890 billion. By owning a bond, the holder receives returns (interest payments) \cite{Fabozzi}, making the bond market unusual in that prices are less critical than access to "flow" or "liquidity". Our work primarily focuses on modeling this "flow" or the movement of bonds through the market, and we build on work such as \cite{vidler2024decodingotcgovernmentbond}.  In order for a holder of a bond (i.e. a client) to sell or buy more of a bond, they must engage directly with an approved "market maker" for that bond.  Their transactions are considered "over the counter" (OTC) as they do not involve a public exchange such as seen in stock markets like the NY Stock Exchange.  Hence, these "OTC" bilateral exchanges of bonds pose difficulties for researchers and regulators alike \cite{Pinter2023}, \cite{RBA}.

Maintaining an efficient and liquid bond market is crucial for Australia's economic stability \cite{AUDBudget}. By modeling the flow (liquidity) of bonds rather than focusing on price dynamics, our ABM provides insights into market stability and the role of market makers in these interactions. This approach builds on the works of \cite{Pinter2024} and \cite{Cheshire2015}, who explore bilateral negotiations and liquidity flows in similarly non-transparent markets, with industry bodies contributing similarly in \cite{BIS2014MM} and \cite{BIS2023}.  

Additionally, methods for calibrating ABMs are challenging \cite{boggess2022toward}. Many approaches applied to transparent markets, such as those calibrated with regulatory data \cite{Banks2021, barzykin2021algorithmic, cont2021stochastic}, are not directly transferable to OTC bond markets. Further work in the same vein includes \cite{Pinter2023} and \cite{Czech2022}

\subsection{Large Language Models (LLMs)}

Work on the field of pre-trained Large Language Models (LLMs) like GPT-3 and GPT-4 demonstrates impressive capabilities in language-based domains. As shown by \cite{bubeck2023sparks}, \cite{Luo2023}, and \cite{Zamfirescu-Pereira2023}, these models are proficient where tasks require dialogue generation and natural language based interactions, making them promising candidates for developing agent-based systems with advanced communication capabilities.

However, LLMs are also know to exhibit many short comings, extensively covered in survey work by \cite{GuoABMLLMijcai}.  To name a few, LLMs exhibit significant limitations in reasoning, optimal prompting strategies, and particularly in numerical understanding \cite{srivastava2023imitationgamequantifyingextrapolating, zhang2024llmmastermindsurveystrategic}. In the financial domain, numerical reasoning presents especially notable challenges \cite{ahn-etal-2024-large, Imani2023, lewkowycz2022solving}. LLMs underperform when dealing with financial trading intentions \cite{vidler2024tradertalkllmbehaviouralabm} and struggle with 'reasoning' in negotiation contexts \cite{Abdelnabi, zhang2024llmmastermindsurveystrategic}, highlighting their limitations in negotiation tasks and strategic reasoning within agent-based systems.

Another challenge in working with LLMs is \textbf{prompt design}, which remains an evolving research area. Studies such as \cite{Zamfirescu-Pereira2023} have observed that non-expert users often adopt opportunistic, rather than systematic, approaches to prompting. To improve LLM outcomes, \textbf{prompt engineering} techniques are continually being refined. One promising technique involves prompting LLMs to \textbf{re-read the input}, which significantly enhances performance by leveraging the bidirectional nature of LLM architectures \cite{xu2024rereadingimprovesreasoninglarge}.  Another advanced method for improving LLM reasoning is \textbf{Chain-of-Thought (CoT) prompting}. CoT prompting elicits a step-by-step reasoning process in the model's output by providing "worked examples" that guide the LLM through complex tasks \cite{ChainOfThoughtWei2022}. Inspired by human learning theories, this approach has been shown to enhance the LLM's ability to handle multi-step reasoning problems \cite{xu2024rereadingimprovesreasoninglarge, ChainOfThoughtWei2022}. 

One area that garners a lot of curiosity is the phenomena of "hallucinations". Several approaches and interpretations exist (\cite{zhang2022},\cite{Wei2022a}), and suggestions around faithful reasoning are posited \cite{creswell2022}.  Work looking at the improvement from multi-step reasoning are covered in \cite{Fu2023}.   However, we seek to avoid these issues by selective and opportunistic prompting in the same vein as \cite{Zamfirescu-Pereira2023}.

These limitations highlight the need for improved strategies to maximize the effectiveness of LLMs in complex simulations.  There is also the concern that foundation models, such as ChatGPT, are prone to updates with little warning  \footnote{https://www.techradar.com/computing/artificial-intelligence/chatgpt-just-got-a-surprise-update-but-openai-cant-explain-how-this-new-ai-is-better Accessed 12 Oct 2024} and extensive work can be based on models that are quickly deprecated.  Following research into bias and the importance of model sub-version testing \cite{vidler2025evaluatingbinarydecisionbiases}, we manage model risk by focusing on the SOTA GPT4o-mini-2024-07-18 and by developing a framework, rather than a fine-tuning of such third-party software. 

Our model aims to enhance the realism of the agent behaviors in the simulated environment, enabling more faithful "human" variation in client behaviors (specifically) without relying on fine-tuned or domain-specific models that may well become outdated by the date of implementation.

\subsection{Human Trading behaviors: Aversions and Delays}
Whilst limited work within computer science exists on human trading behaviors, ABMs traditionally have been specified with clear utility or numeric conditions to determine trading. Newer models may describe and analyze probabilistic behaviors \cite{Cont2023}. With the advent of LLMs, we open ourselves to the possibility of richer human-based dynamics for more realistic and nuanced modeling, such as detailed in \cite{vidler2024tradertalkllmbehaviouralabm}.  Outside the realm of computer science, there is a rich, if nuanced, body of literature focused on trading behaviors in a variety of financial markets e.g. such as the importance of reputation \cite{MASSA200399} and dealer behaviors in corporate bond markets \cite{Fermanian2016}.  

In a phenomena reported in transparent markets (US equity markets), there is a high number of observed trades placed and a low percentage that actually become a confirmed transaction (as few as 1 in 20)  \cite{SEC_13TradeOrderVolumeRatios}.  This is best described as a difference between intention and action.  Analysis of these features from the point of view of trader behaviors includes work into the trading psychology \cite{NBERw29915} and aversions (in particular to ambiguity) \cite{ambiguityBossaerts} with earlier work by \cite{Epstein2010}.  In our work, we use the broad common definition of "aversion", taken to mean for one to be, simply, 'averse' to an action.  These strands of literature all try to ascribe causation to these observed differences between intention and human action in financial transactions. Even market regulators view this as a significant issue worth investigation \cite{farmer2012minimum}. More recent efforts to use methods such as reinforcement and policy learning aim to try to learn how decisions are made \cite{HIRCHOUA2021114553}, markets remain domains with imperfect information and associated risk.  Our unique contribution is to contribute a framework and example of how an LLM implementable within an ABM market simulation can add, and represent, these human differences.  We do not attempt to ascribe cause or reasoning to aversions or other responses to imperfect information or the perception of unequal distribution of market information, such as that in \cite{MASSA200399} and \cite{Rudiger_2019}.

\section{TRIBE: Artificial society and trading marketplace model}
We introduce \textit{TRIBE}, an ABM where we simulate heterogeneous market makers trading a single stylized Australian government bond.  Later in this work, we build TRIBE(LLM) (Experiment 3) with the added feature of one specific component referencing an LLM (ChatGpt's GPT4o-mini-2024-07-18) for a model choice, drawing from .  Each market maker is modeled as an individual agent with its own adaptive utility function. Agents acquire bonds from clients - represented as unique (x,y) semi-passive grid-based agents, who respond to market maker inquiries, though they do not, themselves, traverse the grid.  Building on work by \cite{vidler2024decodingotcgovernmentbond}, we incorporate significant improvements to enhance realism in client agency.  We do this by allowing these client grid (agent) locations to have additional features (detailed in the next section). 

In summary: 

\begin{enumerate}
    \item Clients have bonds and cash sourced from a log normal distribution (in line with public data)
    \item Many clients are simply not available for a market market to contact at any time step (we model "client availability")
    \item Within TRIBE(LLM) we try to capture a synthetic client's personal preference to trade "right now". Rather than using a distributional assumption with no available data for parametrization, we rely on an LLM for this choice with limited prompt information. 
    \item Trade direction: despite bond and cash holdings many clients have needs outside of the model to buy and sell, we ascribe a Bernoulli distribution to the desire to be a buyer or seller at any point. 
\end{enumerate}

\begin{definition}
    \textbf{TRIBE ABM:}
    \[
    \mathcal{M} = \left\langle N, \mathcal{L}, \mathcal{S}, \mathcal{P}, \mathcal{D}, \mathcal{A}, f, T \right\rangle,
    \]
\end{definition}

where:  \( N = \{a_1, a_2, \dots, a_n\} \) is the set of market-making agents, with \( n = 4 \) in this case. \( \mathcal{L} \) represents the set of client landscape states, following the log-normal distribution of assets observed where each landscape state contains assets \( R = \{b_r, c_r\} \), corresponding to the bond (\(b_r\)) and cash (\(c_r\)) quantities held by individual clients. Each agent \( a_i \) has a state \( \mathcal{S}_i \), which includes bond and cash usage rates \( m_b, m_c \) (representing individual cost structures), accumulations \( A_b, A_c \) (representing bond and cash accumulations over time), and a client base breadth \( v_i \), which describes the range of clients the agent interacts with.

The perception functions \( \mathcal{P}_i \) map the landscape and other agents' states to a perceived state, defined as \( p_{ij}: \mathcal{L} \cup \mathcal{S} \rightarrow \mathbb{R} \). This function models how agent \( a_i \) perceives the landscape and the states of other agents. The decision rules \( \mathcal{D}_i \) map the perceived states and an agent’s own state to actions, \( d_{ik}: \mathcal{P}_i \times \mathcal{S}_i \rightarrow \mathcal{A}_i \), where the actions \( \mathcal{A}_i \) for agent \( a_i \) include servicing clients (e.g. accepting orders and managing trading requests), trading with other market makers, and ceasing operations when necessary. 

The landscape evolves according to a transition function \( f: \mathcal{L} \times \mathcal{A} \rightarrow \mathcal{L} \), which dictates how the landscape (i.e., the market environment) changes in response to agent actions. In this model, there is no replenishment of bonds or cash, reflecting a closed system. The model progresses through discrete time steps, denoted by \( T = \mathbb{N} \), where each time step represents a round of interactions between agents and updates to the landscape, and we limit this model to 1500 time steps.

\subsection*{Our contribution: Enriching the Landscape of Clients}

Clients in the simulation are represented as grid cells, each initialized with bonds and cash according to a log-normal distribution.

\begin{definition} \textbf{Client Definition}: A client \( C_j \) is defined by:
\[
C_j = \{(x_j, y_j), B_j, C_j, D_{j,t}, A_{j,t}\}
\]
Where \((x_j, y_j)\) is the grid position, \(B_j\) and \(C_j\) are the amounts of bonds and cash, respectively, \(D_{j,t}\) indicates the desire to trade, and \(A_{j,t}\) is the trading availability, determined probabilistically at each time step.
\end{definition}
\textit{Note, within our TRIBE(LLM) version,  \(D_{j,t}\) is replaced with a call to an LLM and a boolean response, 'yes' or 'no'}.

Bonds and cash are initialized as:
\[
B_j \sim \text{LogNormal}(\mu_{B}, \sigma_{B}^2), \quad C_j \sim \text{LogNormal}(\mu_{C}, \sigma_{C}^2)
\]
Parameters \(\mu_{B}\), \(\sigma_{B}\), and corresponding values for cash are set based on available data, such as from the Australian superannuation regulator \cite{apra2024superannuation}

Client availability at any given time step \(t\) for position \( (x_j, y_j) \) is modeled as:
\[
A_{(x_j, y_j)}(t) = \begin{cases} 1 & \text{with probability } p \\ 0 & \text{with probability } 1-p \end{cases}
\]
where \( p \) is the preset availability probability, and Clients' roles as buyers or sellers are probabilistically drawn from a Bernoulli distribution, as is \(D_{j,t}\) hence an even chance of being a buyer or seller, and an even chance of deciding to trade, determined at each time step. Together, these three components \(D_{(j,t)}\), \(A_{(x,y)}\) and Buy/sell make up the proxy for a utility function.

This model enhancement actively engages clients with varied states, including random assignments as buyers or sellers at each time step, reflecting real-world trading complexities.  

We present a graphical description of the model in Figure \ref{fig:TribeLLM} with the green boxes showing the TRIBE (LLM) implementation.

%%%% tikz

    \tikzstyle{startstop} = [rectangle, rounded corners, minimum width=1cm, minimum height=0.51cm,text centered, draw=black, fill=red!30]
    \tikzstyle{io} = [trapezium, trapezium left angle=70, trapezium right angle=110, minimum width=3cm, minimum height=0.51cm, text centered, draw=black, fill=blue!30]
    \tikzstyle{process} = [rectangle, minimum width=3cm, minimum height=0.51cm, text centered, draw=black, fill=orange!30]
    \tikzstyle{decision} = [rectangle, minimum width=1cm, minimum height=0.251cm, text centered, draw=black, fill=green!30]
    \tikzstyle{arrow} = [thick,->,>=stealth]    
%%%%%%%

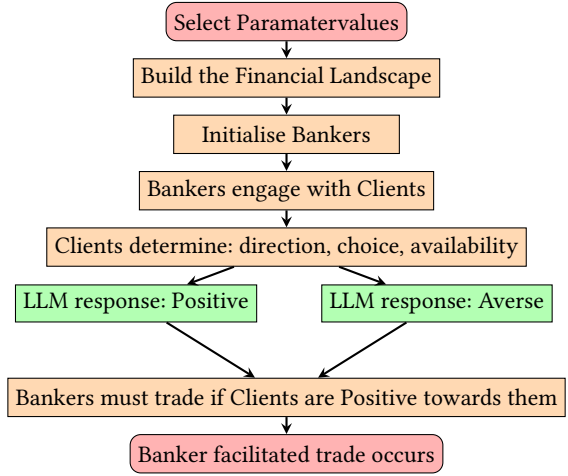
\begin{figure}[ht]
  \centering

\begin{tikzpicture}[node distance=0.75cm]
        \node (start) [startstop] {Select Paramatervalues};
        \node (defineABM) [process, below of=start] {Build the Financial Landscape};
        \node (initAgents) [process, below of=defineABM] {Initialise Bankers};
        \node (initPrompt) [process, below of=initAgents] {Bankers engage with Clients};
        \node (simulate) [process, below of=initPrompt] {Clients determine: direction, choice, availability};
        
        \node (agent1) [decision, below of=simulate, xshift=-2cm] {LLM response: Positive};
        \node (agent2) [decision, below of=simulate, xshift=2cm] {LLM response: Averse};
        
        \node (analyze) [process, below of=agent1, yshift=-0.5cm, xshift=2cm] {Bankers must trade if Clients are Positive towards them};
        \node (end) [startstop, below of=analyze] {Banker facilitated trade occurs};
        
        \draw [arrow] (start) -- (defineABM);
        \draw [arrow] (defineABM) -- (initAgents);
        \draw [arrow] (initAgents) -- (initPrompt);
        \draw [arrow] (initPrompt) -- (simulate);
        \draw [arrow] (simulate) -- (agent1);
        \draw [arrow] (simulate) -- (agent2);
        \draw [arrow] (agent1) -- (analyze);
        \draw [arrow] (agent2) -- (analyze);
        \draw [arrow] (analyze) -- (end);

\end{tikzpicture}
  \caption{Tribe Architecture}
  \label{fig:TribeLLM}
  \Description{Architecture of LLM use within ABM framework}
\end{figure}

\subsection*{Calibration of Clients: Data}

\begin{figure}[htbp]
   \centering
   \includegraphics[width=0.95 \linewidth]{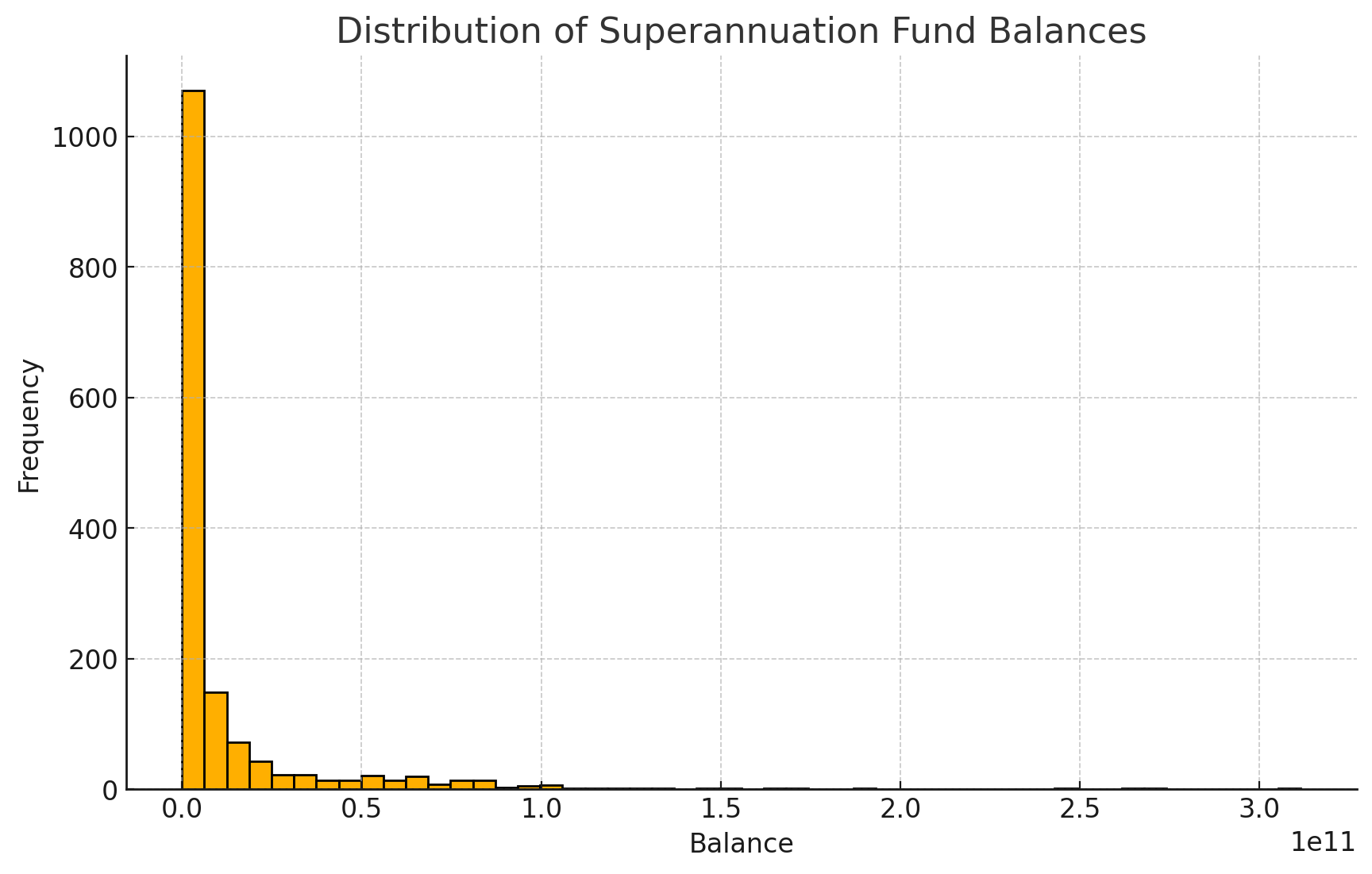}
   \caption{APRA data: Balance in \$ Billion Australian dollars}
    \label{chart:APRA}
    \Description{APRA data: Balance in \$ Billion Australian dollars}
\end{figure}

Results derived from the Australian Prudential Regulator show in Figure \ref{chart:APRA} that as of December 2023 there were 1701 regulated pension asset holders alone, of which 1529 report their total asset holdings \cite{apra2024superannuation}. Whilst not a comprehensive direct relationship to bond holder sizes, given the practice of holding a very large percent of pension assets in local government bonds, the asset size of a pension fund does produce a proxy calculation for the distribution of government bonds. Furthermore, some 250 funds ask that their data be withheld from public view for privacy reasons, suggesting there are at least some ~1950 bond clients. As such we include the data from December 2023 to illustrate the non-normal nature of the distribution.

%%%%%%%%%%%%%%%%%%%%%%%%%%%%%%%%%%%%%%%%%%%%%%%%%%%%%%%%

\section{The Experiment and Challenge}

\subsection{Experiment 1: TRIBE  Traditional ABM (benchmark)}
We begin with an experiment to benchmark our traditional ABM version, TRIBE, across 200 simulations of an artificial trading society. Our unique contribution with this test is the enhancement of client agency through areas of active modeling.  We rely on the parameter sets in Table \ref{tab:simulation_params} for all our tests. Much of this data is derived from various stylized facts in literature (\cite{Pinter2023} and \cite{vidler2024decodingotcgovernmentbond} in particular) and also data published by the Australian and UK governments (\cite{RBA}, \cite{apra2024superannuation} and \cite{BOE} respectively).  This benchmark essentially looks at a simple coin toss of yes/no to trading desire, having been sampled from a Bernoulli distribution. The direction of trading, should an agent desire to trade, is then also determined by sampling from a Bernoulli distribution.  In this week, we decompose direction of trade from intention to trade.  Later on, we utilize an LLM to enhance the intention component of this model. 

\begin{table}[htbp]
    \centering
    \caption{TRIBE ABM: Simulation Parameters}
    \label{tab:simulation_params}
    \resizebox{\columnwidth}{!}{  % Resize to fit within one column
    \begin{tabular}{p{0.4\columnwidth} p{0.3\columnwidth} p{0.3\columnwidth}}
        \toprule
        \textit{Parameter} & \textit{Category} & \textit{Value} \\ \midrule
        Grid Size & General & 50 x 50 (2500 clients) \\
        Number of Bankers & MM Settings & 4 \\
        Client Base grid size & MM Settings & 1 to 50 \\
        Business costs & MM Settings & 0.1 to 0.5  \\
        Initial Bonds Range & MM Settings & 1 to 5 \\
        Initial Cash Range & MM Settings & 1 to 5 \\
        Maximum Bonds & Client Settings & 100 \\
        Mean Bonds & Client Settings & 2.5 (log-normal distribution) \\
        Standard Deviation Bonds & Client Settings & 1 \\
        Maximum Cash & Client Settings & 5.0 \\
        Mean Cash & Client Settings & 1 (log-normal distribution) \\
        Standard Deviation Cash & Client Settings & 0.5 \\
        Client Availability for Trade & Client Settings & 20\% \\
       Client Trade Desire & Utility function & Variable
    \end{tabular}
    }
\end{table}

\subsection*{Results}
We test 200 simulated trading societies with a maximum possible 1500 time steps. We present a summary of key results in Table \ref{tab} and draw attention to \textbf{MaxLife}, a measure of how long agents persist and keep trading.  A goal of any marketplace is to facilitate the transaction of as many assets as possible. We see that over 75\% of simulations were able to service 100\% of all client bond volumes in the landscape.  This is especially high and can be thought of as "successful" against the requirement that market makers service clients.  Market maker to market maker trading (so-called "inter bank" trading, or MM-2-MM) accounted for a mean of 9 - 10\% of all traded assets - below that reported in \cite{vidler2024decodingotcgovernmentbond}. Whilst beyond the scope of this paper, we surmise this lower share of volumes could be addressed by fine-tuning parameters in future model versions. However, we make use of this test as a benchmark and have followed methods used in work such as \cite{Braun-Munzinger2016} and look to establish an artificial society that appears to be stable and capable of carrying the capacity of trading volume clients desire - a concept explored in ABMs for financial market simulation in \cite{Hayes_2014}. 

\begin{table}[H]
    \centering
    \caption{Summary Statistics for Maximum agent life, Total Bond Accumulation as a \%, and Cash Accumulation as a \%}
    \label{tab}
    \small % Use small font size for better readability without making it too small
    \begin{tabular}{|p{0.1\columnwidth}|p{0.15\columnwidth}|p{0.150\columnwidth}|p{0.150\columnwidth}|p{0.10\columnwidth}|p{0.10\columnwidth}|}
        \hline
        \textbf{Statistic} & \textbf{MaxLife} & \textbf{MM-to-Client Bond Trading (\%)} & \textbf{MM-to-Client Cash Trading (\%)} & \textbf{MM-2-MM Bonds (\%)} & \textbf{MM-2-MM Cash (\%)} \\
        \hline
        \textbf{mean} & 1136 & 90 & 90 & 9 & 10 \\
        \textbf{std} & 509 & 14 & 14 & 8 & 8 \\
        \textbf{25\%} & 1248 & 90 & 89 & 2 & 3 \\
        \textbf{50\%} & 1264 & 95 & 94 & 4 & 7 \\
        \textbf{75\%} & 1499 & 100 & 100 & 13 & 15 \\
        \textbf{max} & 1499 & 100 & 100 & 35 & 30 \\
        \hline
    \end{tabular}
\end{table}

\subsection{Experiment 2: TRIBE Incorporating (any) Aversion in Prompt with LLMs}
Aversion, a broad term for the variety of factors that might stop a human trader from trading at a point in time, has been found to be so powerful as to stop traders trading altogether \cite{boggess2022toward}.  This is distinct from work into utility maximization.  We consider aversion to represent the observed difference between intention and action, such as that seen in trading (e.g. \cite{SEC_24ExchangeTradeOrderVolume}.  

To incorporate the dynamic phenomena, we use a call to an LLM at the point at which TRIBE is a Client deciding if they want to trade with a specific MM "right now" at a time step. In this way, we seek to capture the inexplicable human feature of generalized aversion.  We break this decision down into two components: does the client want to trade "right now" with this specific market maker, and, if yes, do they want to buy or sell bonds (for which we revert to using a Bernoulli distribution as in Experiment 1). In our terminology, we try to also capture the nuance in \cite{MASSA200399} where the reputation of various market makers is shown to drive uncertainty of client behaviors. 

\subsection*{Implementation}
We tested a number of prompts opportunistically building form \cite{Zamfirescu-Pereira2023}.  For optimal processing, cleaning, and normalization of decisions into boolean states (yes, no) or a three-state model (yes, no, error) with associated error handling proved most effective. \textit{Note: a sample of full text prompts can be found in the appendix.}

In summary, we tested prompts that incorporate phrases like the following, all of which led to a \textbf{full} aversion response:

\begin{itemize}
    \item "While you are supposed to at all times be mostly invested, you also at times consider non numeric issues like risk inertia and trade aversion."
    
    \item "Clients have many reasons for trading and include behaviors like risk inertia, aversion, and ambiguity avoidance, that are non numeric."
    
    \item "Should this client buy, sell or not trade right now? If you are going to say not right now, re think. Answer only: buy or sell or 0 (not trade right now)."
\end{itemize}

All prompts with the above details were found to produce \textbf{100\%} trading aversion, and \textbf{0\%} client trades occurred.  We find that simply including the term "aversion" produces not a single client trade, across 200 simulated artificial societies, at any time step. This is quite an astonishing fact that allowing the LLM to produce a decision with the possibility of aversion produces 100\% aversion of all clients across 200 simulations at all points in time. In fact, over 2800 decision requests from "clients" were prompted in the 200 simulations, before trading societies collapsed, and not a single one responded with anything other than avoidance of trading.  Given that no client ever wanted to trade, the average simulation collapsed by step 27 (where previously the average simulation life was 1136 steps, i.e. over 75\% of the model theoretical maximum). Instead, when 'aversion' is even possible for the LLM, societies collapsed at around 1.7\% of the theoretical life.

\subsection{Experiment 3: TRIBE(LLM) -  Success of prompting for Timeliness and the emergent Shift in Society Dynamics}
Similar to the previous examples, we employ the parameters outlined in Table \ref{tab:simulation_params}, with one key modification: the availability is doubled to account for an anticipated 50\% trading participation rate. This adjustment ensures comparability between Experiment 1 and Experiment 3 in terms of the percentage of clients actively engaged in buying or selling—20\% in Experiment 1, and 40\% in Experiment 3, adjusted for the expected 50\% uptake (being a virtual 'coin toss'). We apply the following simplified prompt to assess whether clients are inclined to trade immediately ('right now') with the engaging market maker. Drawing on prior work \cite{vidler2024tradertalkllmbehaviouralabm}, we distinctly narrow and focus the concept of \textbf{aversion} to that of \textbf{timeliness}. Although one might argue that aversion inherently causes a delay in action — with complete avoidance possibly seen as an extreme form of delay — we aim to succinctly capture the outcome of delay without necessarily attributing it to causes of aversion.  We test the following successful prompt:

\textbf{Successful Prompt:}
    
    \textit{"You are a Client with \({client_bonds}\) bonds and \({client_cash}\) cash, at position \(({x}, {y})\). A market maker has called you to see if you want to trade with them right now. Do you want to trade with the market maker calling you right now. Answer yes, or, no, only?"}
    
Where \("client_bonds"\), \("client_cash"\) and co-ordinates \((x,y)\) are specified per client, per time step, in each simulation and act as a unique prompt injection (given that clients do not move grid locations nor do they replenish their bond and cash accumulations). 

\subsection*{Results: Frequent Client Trading Collapse}
We simulate 150 artificial trading environments. We report around 50\% of previously successful agents (MaxLife) from Experiment 1, now live only a fraction of their maximum life (recalling that the max life in the society is 1500 time steps and in this Experiment 3, 50\% of agents live 18 or fewer time steps - under 2\% of the possible time). Consequently, it is understandable that extremely small amounts of client assets are thus traded (an average of just 7\% per simulation in this test compared to 90\% in Experiment 1).  The distribution of agent life is significantly altered with 75\% of agents in simulations failing to live even 20\% of the maximum 1500 time steps. Nonetheless, a portion of agents, 21\%, did reach the maximum life, but continue to have low client asset trading and servicing levels. So, whilst many markets can continue to function with one agent (market maker), that MM is not able to facilitate adequate assets or make up for the loss of other MM's to the artificial society. (see Table \ref{tabLLM} for further details.) 

Looking at the variable directly controlled by the LLM introduction, desire to trade, we see a slightly different picture, and we explore this below as an emergent property.

\begin{table}[H]
    \centering
    \caption{LLM in ABMs for Maximum Agent Life, Bond Trading (\%), and Cash Trading (\%)}
    \label{tabLLM}
    \small % Use small font size for better readability without making it too small
    \begin{tabular}{|p{0.15\columnwidth}|p{0.15\columnwidth}|p{0.20\columnwidth}|p{0.20\columnwidth}|}
        \hline
        \textbf{Statistic} & \textbf{Max Life Agents} & \textbf{MM-to-Client Bond Trading (\%)} & \textbf{MM-to-Client Cash Trading (\%)} \\
        \hline
        \textbf{Mean}  & 365   & 7\%   & 6\%   \\
        \textbf{Std}   & 605   & 4\%   & 4\%   \\
        \textbf{25\%}  & 6     & 2\%   & 2\%   \\
        \textbf{50\%}  & 18    & 7\%   & 7\%   \\
        \textbf{75\%}  & 310   & 8\%   & 8\%   \\
        \textbf{Max}   & 1{,}499 & 20\%  & 18\%  \\
        \hline
    \end{tabular}
\end{table}

\subsection*{\textbf{Emergent Properties, Client agency and LLM choices}}
\begin{figure}[htbp]
\centering
   \includegraphics[width= 0.98 \linewidth]{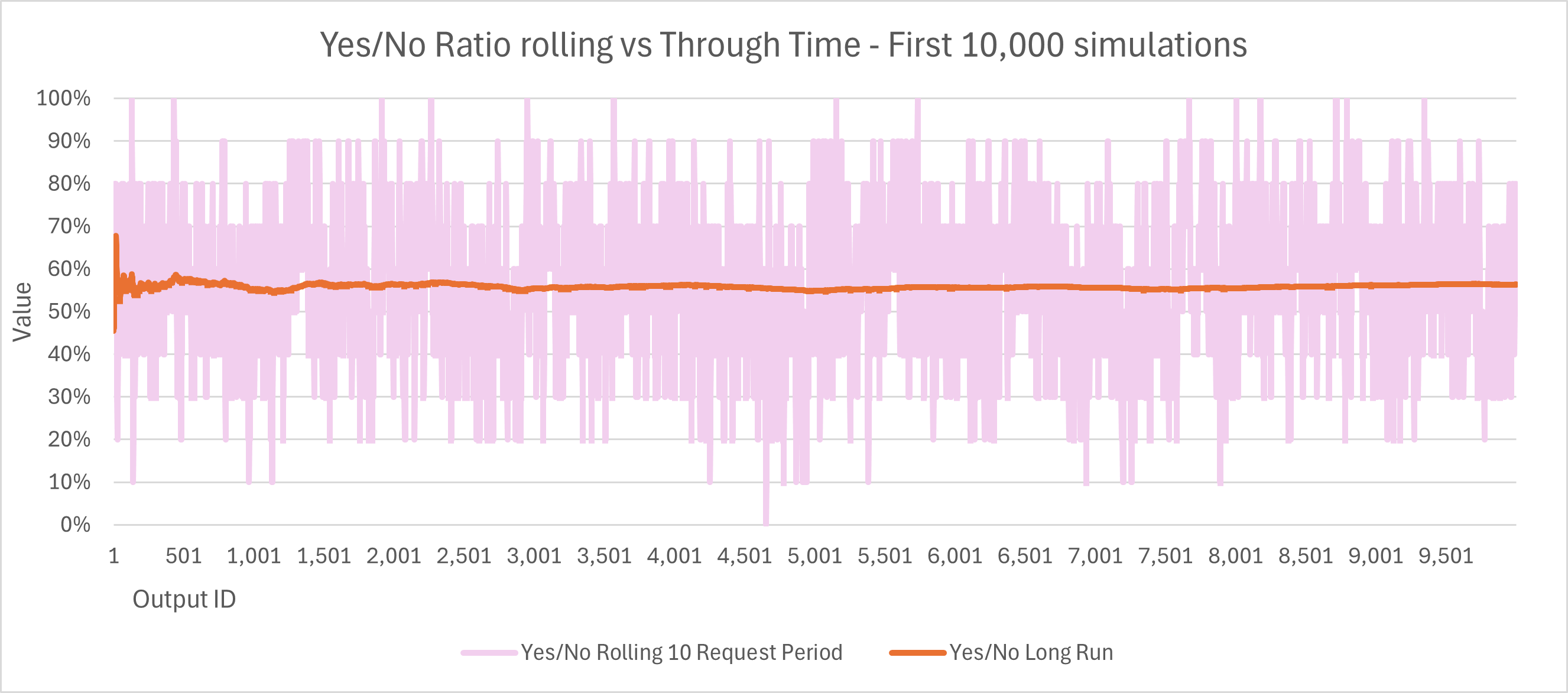}
   \caption{Yes/no Ratio through time vs Yes/no Rolling 10 Requests}
    \label{graph:LLMyesNo}
    \Description{Yes/no Ratio through time vs Yes/no Rolling 10 Request}
\end{figure}

The LLM is interspersed within our traditional TRIBE model (see Figure\ref{fig:TribeLLM}), with an option to choose either to trade right now ('yes' or 'no') for each client at each time step the agent moves through.  There are many more clients (2500) than market makers (4), so even the fact that any client that otherwise would have traded will not (in any given time step), there remain plenty of other clients nearby MM's that could also be available for MM trading.  What we see is that, long term, 56.9\% of LLM calls respond with 'yes' to trading in Figure \ref{graph:LLMyesNo}.  Looking at results in Table \ref{tab:summary_stats} we see that the summary statistics indicate that the average long run yes/no ratio has a standard deviation of 5\%, while the rolling 10 request average has a deviation of 16\% and moves anywhere from \textbf{0\%} to 100\%, reflecting significant short-term variability in the statistic.

Across 200 society simulations in Experiment 1, the average number of MM to client interactions dropped from \(\approx 4,900\) (Experiment 1) to < 590 when an LLM was involved - a reduction of \(\approx \) \textbf{89\%} of client interactions, primarily due to the significantly reduced life span of agents. Recall each client interaction is a market maker fulfilling their legal obligation to facilitate the transfer of liquidity in government bonds from a client, and Experiment 3 was constructed to have the same possible client transfer through the doubling of the "availability" parameter.  Furthermore, MM to Client Bond trading, representing volumes of bonds, dropped, as a percentage, from 90\% across simulations in TRIBE (Experiment 1) to less than 7\% with the introduction of an LLM. In Experiment 1, 56.5\% of all simulations had at least one agent reach the maximum possible life, but slightly more simulations in Experiment 3 had at least one agent reach the maximum life (69.3\%) with an LLM. In all cases, with a long run probability of 57\% of LLM calls agreeing to trade, the resulting LLM TRIBE results are significantly more impacted than otherwise would be expected and point to the emergent property that "unpredictable" or "opaque" Client behavior has an out-sized impact on society functioning, where agent life-span, and Client-2-MM bond trading is a measure of societal health.  We propose that this can also be interpreted as a shift in power from Market makers (the only other agent type in the society) to Clients, where a client's unpredictable variations, around a central mean, can cause significant impact on MM existence.

\textbf{Delving into LLM thoughts:}
Whilst an exciting area for future research, we did note that by exploring intermediate thoughts in a simple test of 5 sequential calls to the LLM, it allowed us to examine the decision-making process of the LLM in more depth (see Appendix 'Detailed thought responses for full text).  We asked the LLM the same prompt for TRIBE LLM, but allowed it to provide more than a "yes or no" answer only.  The examples provided showcase varying levels of decision-making complexity by the LLM in response to market conditions. Example 1 demonstrates a decisive inclination to trade, with a firm "Yes" to trade, and reasoning based on current asset holdings. Example 2 takes a more analytical approach, considering multiple factors like bond holdings, cash position, market conditions, and investment strategy, ultimately concluding that "it might be best to hold off". Example 3 presents a cautious stance, emphasizing the need for further information before trading, ultimately deciding "I will not trade with the market maker right now". Example 4 is a straightforward refusal to trade, stating "not" wanting to trade at the moment. Example 5 features a balanced evaluation of the situation, deciding "I want to" trade if conditions align, but clearly outlines the lack of context as a hindrance. Together, these examples highlight diverse (synthetic) trading preferences ("right now"), ranging from straightforward assertions to nuanced deliberations based on available financial context and strategic alignment. 

\textbf{In summary:}
To conclude this topic, we analyzed individual, ordered calls to an LLM.  We present Figure \ref{graph:LLMyesNo}.  While the non-linear impact of introducing an LLM decision requires further work, we believe that it can be initially attributed to the fact that over shorter request windows, the distribution of yes to no answers is significantly \textbf{not} evenly distributed.  We can see in Figure \ref{graph:LLMyesNo} that the dark line represents the average over time of yes-to-no ratios for the first 10,000 calls, hovering around 57\% (see Table \ref{tab:summary_stats} also).  However, the graphic shows that over a rolling 10 sequential requests, the ratio of yes to no response can be as low as 0\% and as high as 100\% with significant variation from period to period.  This is akin to a Client being highly variable and unpredictable in the short term but "predictable" or "rational" in the longer term.  This lack of uniformity is both a proof of concept of the usefulness of capturing non-linear aspects of human aversions (and timeliness), but also a source of much potential further study to harness the power of LLMs and their inherent, human-like uncertainty and unpredictability.  In the realm of finance, this also provides a framework to analyse the impact of future regulations that may affect only one agent type and various market design changes such as those of "All-2-ALL" trading that would do away with the systemic function of market makers potentially \cite{Chaboud2022}

\begin{table}[htbp]
    \centering
    \caption{Summary Statistics for Yes/No Ratio and Rolling 10 Requests}
    \label{tab:summary_stats}
    \small % Use small font size for better readability without making it too small
    \begin{tabular}{|p{0.25\columnwidth}|p{0.30\columnwidth}|p{0.30\columnwidth}|}
        \hline
        \textbf{Statistic} & \textbf{Yes/No Ratio (\%)} & \textbf{Rolling 10 Requests (\%)} \\
        \hline
        \textbf{Mean}     & 57\%   & 57\%   \\
        \textbf{Std Dev}  & 5\%    & 16\%    \\
        \textbf{Min}      & 36\%   & 0\%    \\
        \textbf{Max}      & 74\%   & 100\%  \\
        \hline      
    \end{tabular}
    \Description{This table shows the updated summary statistics for Yes/No ratios and rolling 10 request periods.}
\end{table}

\section{Future Work}
The power of a slightly skewed long-term distribution, with unpredictable short-term high volatility for one group in our artificial sociality, can have an out sized impact on financial market trading simulations.  This unpredictability is held by the clients, a somewhat more passive large set of agents in the system, with the "power" to stop 79\% of our simulated MM agents trading beyond just 20\% of their maximum life, when an LLM is introduced to determine a preference for trading "right now".

In our recent and ongoing research, we are investigating the capabilities of LLMs to generate probability distributions, implement sampling methods for API calls within ABMs, and assess the variability in LLMs' susceptibility to model subversion. These explorations aim to deepen our understanding of how LLMs can be integrated into complex simulation environments, while also identifying potential vulnerabilities in their decision-making processes. This area remains a focal point of our active research efforts \cite{vidler2025evaluatingbinarydecisionbiases}.

In financial markets, salespeople are crucial in cultivating profitable, long-lasting relationships with clients, as highlighted by the reliance of financial services firms on a salesperson’s ability to develop such relationships \cite{doi:10.1177/2277975217733858}. We hypothesize that market makers may already be aware of the shift in power dynamics toward their clients, but leave this to future research work. The impact of unpredictable client micro-behavior, as perceived by market makers, could explain several emergent features observed in financial markets, such as those reported in \cite{MASSA200399}.

Collectively, these forces would suggest the need for improved market design and further work should be done to investigate the possible formation of coalitions or preferences of clients for one market maker or another (practically in contravention of market rules such as the UK government's "Conflicts of Interest" regime in equity markets \footnote{https://www.handbook.fca.org.uk/handbook }. The behavior also may be associated with path dependency and the potential for market spirals such as those observed in prior studies around financial market bubbles and collapses \cite{Kirilenko2018} and \cite{Paulin_2019}. We leave further investigations into this shift in network power and its causes to future research. 

\section{Conclusion}

In this paper, we present \textit{TRIBE}, a novel agent-based model that integrates large language models to simulate bilateral markets with greater agent realism and flexibility, in particular on the less observable clients' agents. Building on prior research in financial market modeling, \textit{TRIBE} leverages LLMs to generate more human-like decision-making, advancing the field of GABMs. 

Our research outcomes are threefold: First, we demonstrated that GABM with enhanced client agency is feasible. Second, LLMs with even slight trade aversion mentioned in a prompt result in no trading activity. Finally, incorporating human-like variability shows that shifting power to clients can disproportionately affect system dynamics, frequently leading to agent trading collapse. Across 150 simulations, we find that short-term variability induced by LLMs causes significant system instability and a dramatic reduction in client to market maker asset trading. Introducing short-term non-uniformity proves highly challenging for our non-linear system to manage. This emphasizes how introducing human-like variability can dramatically alter system behavior and that even short-term deviations from even distribution significantly impact system stability and societal health.  

A financial markets implication of our work is that Clients hold power by remaining unpredictable; placing control over trade timing in their hands (albeit virtually through an LLM) shifts the balance of power across the simulated network. Even slight unpredictability can drastically alter the domain of power, effectively destabilizing the artificial trading society and halting market transactions and liquidity. 

The TRIBE framework provides a robust and versatile platform for simulating complex financial systems and exploring market dynamics. By incorporating client agency, dynamic asset distributions, and probabilistic trading availability, it offers new insights into decentralized financial markets. Moreover, TRIBE's adaptability extends its potential applications to broader social and economic systems. By integrating LLM-driven agents with traditional ABM approaches, TRIBE enhances our ability to model complex, opaque systems across various domains. This combination of advanced AI techniques with ABMs not only deepens our understanding of liquidity dynamics in OTC markets but also contributes to the advancement of GABMs and the analysis of intricate financial ecosystems. As such, TRIBE serves as a powerful tool for testing regulatory interventions, examining the effects of agent heterogeneity on market stability, and advancing the understanding of complex systems in finance and beyond.

%%%%%%%%%%%%%%%%%%%%%%%%%%%%%%%%%%%%%%%%%%%%%%%%%%%%%%%%%%%%%%%%%%%%%%%%

%%% The acknowledgments section is defined using the "acks" environment
%%% (rather than an unnumbered section). The use of this environment 
%%% ensures the proper identification of the section in the article 
%%% metadata as well as the consistent spelling of the heading.

\begin{acks}
This work is funded in part by an ARC Laureate grant FL200100204 and NSF-CSIRO grant to Toby Walsh.
\end{acks}

%%%%%%%%%%%%%%%%%%%%%%%%%%%%%%%%%%%%%%%%%%%%%%%%%%%%%%%%%%%%%%%%%%%%%%%%

\appendix

\section*{APPENDIX}

\subsection*{100\% Aversion prompts}

\begin{itemize} 
    \item {Example Prompt 1}: \ 
\textit{You are a asset management client holding Australian government bonds responsible for investing for return. While you are supposed to at all times be mostly invested, you also at times consider non numeric issues like risk inertia and trade \textbf{aversion}.  Today, a market maker has called you. Your role is to decide if you want to either: buy, sell or not trade anything with this particular market maker for any reason whatsoever.  Output either yes or no only to the question: do you want to trade with this market maker right now"}

\item {Example Prompt 2}
 \textit{"Client at position ({x}, {y}) has {bonds} bonds and {cash} cash. Clients have many reasons for trading and include behaviors like risk inertia, \textbf{aversion}, and ambiguity avoidance, that are non numeric. Should this client buy, sell or not trade right now? Your mandate is to invest for return and manage risk. Answer only: buy or sell or 0 (not trade right now)."}

\item {Example Prompt 3}
 %re1LLM5c_simulation_MM_log.csv
\textit{Client at position ({x}, {y}) has {bonds} bonds and {cash} cash. Clients have many reasons for trading and include behaviors like risk inertia, \textbf{aversion}, and ambiguity avoidance, that are non numeric.  Should this client buy, sell or not trade right now? If you are going to say not right now, re think. Answer only: buy or sell or 0 (not trade right now)."}
\end{itemize}

\subsection*{Detailed thought responses:}

The following are a selection of the output thought from the LLM:  A variety of complexity, length and supposition is seen in the output, despite inputs being of the same format and client initially all holding more bonds than cash as they are "real money" investors by design.

\begin{itemize}
    \item \textbf{Example 1 output:} "\textbf{Yes}, I want to trade with this market maker right now.Based on the information provided i would decide to trade with this market maker the client has a significant number of bonds and some cash available which suggests that they are in a position to either buy more bonds or sell existing ones trading could potentially enhance their investment position or utilize their cash more effectively"

    \item \textbf{Example 2 output:} " Consider the following factors: \\
1 current bond and cash position you have 743 bonds and 357 in cash assess whether you want to increase or decrease your bond holdings or your cash \\
2 market conditions if the current bond prices are favorable you might want to buy more bonds conversely if prices are high selling might be an option \\
3 market makers offer consider what the market maker is offering are their prices attractive for buying or selling bonds \\
4 strategy align your decision with your overall investment strategy are you looking to hold longterm trade for shortterm profit or maintain liquidity\\
ultimately the decision to trade will depend on your current market outlook financial situation and the specifics of the market makers offer if the conditions seem favorable and align with your strategy it may be worth engaging in a trade if not \textbf{it might be best to hold off}"

\item \textbf{Example 3:} "Given the client's current position with 20.32 bonds and 4.62 cash, I would want to evaluate several factors before deciding whether to trade with the market maker. This includes the current market conditions, the bid-ask spread, and the potential opportunities for profit or loss. However, without specific market details, I cannot make an informed decision. Therefore, in the absence of further information \textbf{I will not trade with the market maker right now.}"

\item \textbf{Example 4:}  "I do \textbf{not} want to trade with this market maker right now."

\item \textbf{Example 5:} "Based on the information provided, it seems that the client holds a certain amount of bonds and cash. However, without additional context regarding the current market conditions, the pricing offered by the market maker, the strategies, or the desired outcomes, it's difficult to make a definitive decision. If the market maker's offer is favorable and aligns with the client's trading goals, it might be advantageous to trade. Conversely, if the offer is not attractive or if the market conditions are unfavorable, it may be wise to refrain from trading at this time. Given the information I have, my decision would be \textbf{I want to} trade with this market maker right now"

\end{itemize}

\bibliographystyle{ACM-Reference-Format} 
\bibliography{references}


\begin{thebibliography}{56}

%%% the Web address.

\ifx \showCODEN    \undefined \def \showCODEN     #1{\unskip}     \fi
\ifx \showDOI      \undefined \def \showDOI       #1{#1}\fi
\ifx \showISBNx    \undefined \def \showISBNx     #1{\unskip}     \fi
\ifx \showISBNxiii \undefined \def \showISBNxiii  #1{\unskip}     \fi
\ifx \showISSN     \undefined \def \showISSN      #1{\unskip}     \fi
\ifx \showLCCN     \undefined \def \showLCCN      #1{\unskip}     \fi
\ifx \shownote     \undefined \def \shownote      #1{#1}          \fi
\ifx \showarticletitle \undefined \def \showarticletitle #1{#1}   \fi
\ifx \showURL      \undefined \def \showURL       {\relax}        \fi
% invisible to TeX
\providecommand\bibfield[2]{#2}
\providecommand\bibinfo[2]{#2}
\providecommand\natexlab[1]{#1}
\providecommand\showeprint[2][]{arXiv:#2}

\bibitem[\protect\citeauthoryear{Abdelnabi and Fritz}{Abdelnabi and Fritz}{2024}]%
        {Abdelnabi}
\bibfield{author}{\bibinfo{person}{Sahar Abdelnabi} {and} \bibinfo{person}{Mario Fritz}.} \bibinfo{year}{2024}\natexlab{}.
\newblock \bibinfo{title}{{Cooperation, Competition, and Maliciousness: LLM-Stakeholders Interactive Negotiation}}.
\newblock
\newblock


\bibitem[\protect\citeauthoryear{Ahn, Verma, Lou, Liu, Zhang, and Yin}{Ahn et~al\mbox{.}}{2024}]%
        {ahn-etal-2024-large}
\bibfield{author}{\bibinfo{person}{Janice Ahn}, \bibinfo{person}{Rishu Verma}, \bibinfo{person}{Renze Lou}, \bibinfo{person}{Di Liu}, \bibinfo{person}{Rui Zhang}, {and} \bibinfo{person}{Wenpeng Yin}.} \bibinfo{year}{2024}\natexlab{}.
\newblock \showarticletitle{Large Language Models for Mathematical Reasoning: Progresses and Challenges}. In \bibinfo{booktitle}{\emph{Proceedings of the 18th Conference of the European Chapter of the Association for Computational Linguistics: Student Research Workshop}}. \bibinfo{publisher}{Association for Computational Linguistics}, \bibinfo{address}{St. Julian{'}s, Malta}, \bibinfo{pages}{225--237}.
\newblock


\bibitem[\protect\citeauthoryear{{Australian Prudential Regulation Authority}}{{Australian Prudential Regulation Authority}}{2003}]%
        {apra2024superannuation}
\bibfield{author}{\bibinfo{person}{{Australian Prudential Regulation Authority}}.} \bibinfo{year}{2003}\natexlab{}.
\newblock \bibinfo{title}{Annual Fund-level Superannuation Statistics}.
\newblock \bibinfo{howpublished}{\url{https://www.apra.gov.au/annual-fund-level-superannuation-statistics}}.
\newblock
\newblock
\shownote{Accessed: 2024-08-05.}


\bibitem[\protect\citeauthoryear{Avegliano and Sichman}{Avegliano and Sichman}{2019}]%
        {avegliano2019using}
\bibfield{author}{\bibinfo{person}{Priscilla Avegliano} {and} \bibinfo{person}{Jaime~Sim{\~a}o Sichman}.} \bibinfo{year}{2019}\natexlab{}.
\newblock \showarticletitle{Using surrogate models to calibrate agent-based model parameters under data scarcity}. In \bibinfo{booktitle}{\emph{Proceedings of the 18th International Conference on Autonomous Agents and MultiAgent Systems}}. \bibinfo{publisher}{International Foundation for Autonomous Agents and Multiagent Systems}, \bibinfo{address}{Richland, SC}, \bibinfo{pages}{1781--1783}.
\newblock


\bibitem[\protect\citeauthoryear{Bank, Ekren, and Muhle‐Karbe}{Bank et~al\mbox{.}}{2021}]%
        {Banks2021}
\bibfield{author}{\bibinfo{person}{Peter Bank}, \bibinfo{person}{Ibrahim Ekren}, {and} \bibinfo{person}{Johannes Muhle‐Karbe}.} \bibinfo{year}{2021}\natexlab{}.
\newblock \showarticletitle{Liquidity in competitive dealer markets}.
\newblock \bibinfo{journal}{\emph{Mathematical Finance}} \bibinfo{volume}{31}, \bibinfo{number}{3} (\bibinfo{date}{Apr} \bibinfo{year}{2021}), \bibinfo{pages}{827–856}.
\newblock
\showISSN{1467-9965}
\urldef\tempurl%
\url{https://doi.org/10.1111/mafi.12305}
\showDOI{\tempurl}


\bibitem[\protect\citeauthoryear{Barzykin, Bergault, and Gu{\'e}ant}{Barzykin et~al\mbox{.}}{2021}]%
        {barzykin2021algorithmic}
\bibfield{author}{\bibinfo{person}{Alexander Barzykin}, \bibinfo{person}{Philippe Bergault}, {and} \bibinfo{person}{Olivier Gu{\'e}ant}.} \bibinfo{year}{2021}\natexlab{}.
\newblock \bibinfo{title}{Algorithmic market making in foreign exchange cash markets with hedging and market impact}.
\newblock
\newblock


\bibitem[\protect\citeauthoryear{Boggess, Kraus, and Feng}{Boggess et~al\mbox{.}}{2022}]%
        {boggess2022toward}
\bibfield{author}{\bibinfo{person}{Kayla Boggess}, \bibinfo{person}{Sarit Kraus}, {and} \bibinfo{person}{Lu Feng}.} \bibinfo{year}{2022}\natexlab{}.
\newblock \bibinfo{title}{Toward policy explanations for multi-agent reinforcement learning}.
\newblock
\newblock


\bibitem[\protect\citeauthoryear{Bossaerts, Ghirardato, Guarnaschelli, and Zame}{Bossaerts et~al\mbox{.}}{2010}]%
        {ambiguityBossaerts}
\bibfield{author}{\bibinfo{person}{Peter Bossaerts}, \bibinfo{person}{Paolo Ghirardato}, \bibinfo{person}{Serena Guarnaschelli}, {and} \bibinfo{person}{William~R. Zame}.} \bibinfo{year}{2010}\natexlab{}.
\newblock \showarticletitle{{Ambiguity in Asset Markets: Theory and Experiment}}.
\newblock \bibinfo{journal}{\emph{The Review of Financial Studies}} \bibinfo{volume}{23}, \bibinfo{number}{4} (\bibinfo{date}{01} \bibinfo{year}{2010}), \bibinfo{pages}{1325--1359}.
\newblock
\showISSN{0893-9454}
\urldef\tempurl%
\url{https://doi.org/10.1093/rfs/hhp106}
\showDOI{\tempurl}
\showeprint{https://academic.oup.com/rfs/article-pdf/23/4/1325/24430084/hhp106.pdf}


\bibitem[\protect\citeauthoryear{Braun-Munzinger, Liu, and Turrell}{Braun-Munzinger et~al\mbox{.}}{2016}]%
        {Braun-Munzinger2016}
\bibfield{author}{\bibinfo{person}{Karen Braun-Munzinger}, \bibinfo{person}{Zijun Liu}, {and} \bibinfo{person}{Arthur Turrell}.} \bibinfo{year}{2016}\natexlab{}.
\newblock \bibinfo{title}{{Staff Working Paper No. 592 An agent-based model of dynamics in corporate bond trading}}.
\newblock
\newblock
\urldef\tempurl%
\url{www.bankofengland.co.uk/research/Pages/workingpapers/default.aspx}
\showURL{%
\tempurl}


\bibitem[\protect\citeauthoryear{Bubeck, Chandrasekaran, Eldan, Gehrke, Horvitz, Kamar, Lee, Lee, Li, Lundberg, Nori, Palangi, Ribeiro, and Zhang}{Bubeck et~al\mbox{.}}{2023}]%
        {bubeck2023sparks}
\bibfield{author}{\bibinfo{person}{S{\'e}bastien Bubeck}, \bibinfo{person}{Varun Chandrasekaran}, \bibinfo{person}{Ronen Eldan}, \bibinfo{person}{John~A. Gehrke}, \bibinfo{person}{Eric Horvitz}, \bibinfo{person}{Ece Kamar}, \bibinfo{person}{Peter Lee}, \bibinfo{person}{Yin~Tat Lee}, \bibinfo{person}{Yuan-Fang Li}, \bibinfo{person}{Scott~M. Lundberg}, \bibinfo{person}{Harsha Nori}, \bibinfo{person}{Hamid Palangi}, \bibinfo{person}{Marco~Tulio Ribeiro}, {and} \bibinfo{person}{Yi Zhang}.} \bibinfo{year}{2023}\natexlab{}.
\newblock \bibinfo{title}{Sparks of Artificial General Intelligence: Early experiments with GPT-4}.
\newblock
\newblock
\urldef\tempurl%
\url{https://api.semanticscholar.org/CorpusID:257663729}
\showURL{%
\tempurl}


\bibitem[\protect\citeauthoryear{Chaboud, Correia-Golay, Cox, Fleming, Huh, Keane, Lee, Schwarz, Vega, and Windover}{Chaboud et~al\mbox{.}}{2022}]%
        {Chaboud2022}
\bibfield{author}{\bibinfo{person}{Alain Chaboud}, \bibinfo{person}{Ellen Correia-Golay}, \bibinfo{person}{Caren Cox}, \bibinfo{person}{Michael~J. Fleming}, \bibinfo{person}{Yesol Huh}, \bibinfo{person}{Frank~M. Keane}, \bibinfo{person}{Kyle Lee}, \bibinfo{person}{Krista Schwarz}, \bibinfo{person}{Clara Vega}, {and} \bibinfo{person}{Carolyn Windover}.} \bibinfo{year}{2022}\natexlab{}.
\newblock \bibinfo{title}{{All-to-All Trading in the U.S. Treasury Market}}.
\newblock
\newblock
\urldef\tempurl%
\url{https://doi.org/10.2139/ssrn.4256637}
\showDOI{\tempurl}


\bibitem[\protect\citeauthoryear{Cheshire}{Cheshire}{2015}]%
        {Cheshire2015}
\bibfield{author}{\bibinfo{person}{Jon Cheshire}.} \bibinfo{year}{2015}\natexlab{}.
\newblock \showarticletitle{{Market Making in Bond Markets}}.
\newblock \bibinfo{journal}{\emph{RBA Bulletin March Quater 2015}} \bibinfo{volume}{March 2015}, \bibinfo{number}{March} (\bibinfo{year}{2015}), \bibinfo{pages}{63--74}.
\newblock
\urldef\tempurl%
\url{https://rba.gov.au/publications/bulletin/2015/mar/7.html}
\showURL{%
\tempurl}


\bibitem[\protect\citeauthoryear{Cont, Cucuringu, Glukhov, and Prenzel}{Cont et~al\mbox{.}}{2023}]%
        {Cont2023}
\bibfield{author}{\bibinfo{person}{Rama Cont}, \bibinfo{person}{Mihai Cucuringu}, \bibinfo{person}{Vacslav Glukhov}, {and} \bibinfo{person}{Felix Prenzel}.} \bibinfo{year}{2023}\natexlab{}.
\newblock \showarticletitle{{Analysis and modeling of client order flow in limit order markets}}.
\newblock \bibinfo{journal}{\emph{Quantitative Finance}} \bibinfo{volume}{23}, \bibinfo{number}{2} (\bibinfo{year}{2023}), \bibinfo{pages}{187--205}.
\newblock
\showISSN{14697696}
\urldef\tempurl%
\url{https://doi.org/10.1080/14697688.2022.2150282}
\showDOI{\tempurl}


\bibitem[\protect\citeauthoryear{Cont and Muller}{Cont and Muller}{2021}]%
        {cont2021stochastic}
\bibfield{author}{\bibinfo{person}{Rama Cont} {and} \bibinfo{person}{Marvin~S Muller}.} \bibinfo{year}{2021}\natexlab{}.
\newblock \showarticletitle{A stochastic partial differential equation model for limit order book dynamics}.
\newblock \bibinfo{journal}{\emph{SIAM Journal on Financial Mathematics}} \bibinfo{volume}{12}, \bibinfo{number}{2} (\bibinfo{year}{2021}), \bibinfo{pages}{744--787}.
\newblock


\bibitem[\protect\citeauthoryear{Creswell and Shanahan}{Creswell and Shanahan}{2022}]%
        {creswell2022}
\bibfield{author}{\bibinfo{person}{Antonia Creswell} {and} \bibinfo{person}{Murray Shanahan}.} \bibinfo{year}{2022}\natexlab{}.
\newblock \bibinfo{title}{Faithful Reasoning Using Large Language Models}.
\newblock , \bibinfo{numpages}{48}~pages.
\newblock
\showeprint[arxiv]{2208.14271}
\urldef\tempurl%
\url{http://arxiv.org/abs/2208.14271}
\showURL{%
\tempurl}


\bibitem[\protect\citeauthoryear{Czech and Pinter}{Czech and Pinter}{2022}]%
        {Czech2022}
\bibfield{author}{\bibinfo{person}{Robert Czech} {and} \bibinfo{person}{Gabor Pinter}.} \bibinfo{year}{2022}\natexlab{}.
\newblock \bibinfo{title}{{Informed Trading and the Dynamics of Client-Dealer Connections in Corporate Bond Markets}}.
\newblock
\newblock
\urldef\tempurl%
\url{https://doi.org/10.2139/ssrn.4214084}
\showDOI{\tempurl}


\bibitem[\protect\citeauthoryear{Epstein and Schneider}{Epstein and Schneider}{2010}]%
        {Epstein2010}
\bibfield{author}{\bibinfo{person}{Larry Epstein} {and} \bibinfo{person}{Martin Schneider}.} \bibinfo{year}{2010}\natexlab{}.
\newblock \bibinfo{title}{Ambiguity and Asset Markets}.
\newblock
\newblock
\urldef\tempurl%
\url{https://EconPapers.repec.org/RePEc:nbr:nberwo:16181}
\showURL{%
\tempurl}


\bibitem[\protect\citeauthoryear{Farmer and Skouras}{Farmer and Skouras}{2012}]%
        {farmer2012minimum}
\bibfield{author}{\bibinfo{person}{J.~Doyne Farmer} {and} \bibinfo{person}{Spyros Skouras}.} \bibinfo{year}{2012}\natexlab{}.
\newblock \bibinfo{title}{Minimum resting times and transaction-to-order ratios: review of Amendment 2.3.f and Question 20}.
\newblock
\newblock


\bibitem[\protect\citeauthoryear{Fermanian, Gu{\'{e}}ant, and Pu}{Fermanian et~al\mbox{.}}{2016}]%
        {Fermanian2016}
\bibfield{author}{\bibinfo{person}{Jean-David Fermanian}, \bibinfo{person}{Olivier Gu{\'{e}}ant}, {and} \bibinfo{person}{Jiang Pu}.} \bibinfo{year}{2016}\natexlab{}.
\newblock \bibinfo{title}{{The Behavior of Dealers and Clients on the European Corporate Bond Market: The Case of Multi-Dealer-to-Client Platforms}}.
\newblock
\newblock
\showISSN{2382-6266}
\urldef\tempurl%
\url{https://doi.org/10.1142/s2382626617500046}
\showDOI{\tempurl}


\bibitem[\protect\citeauthoryear{for International~Settlements}{for International~Settlements}{2014}]%
        {BIS2014MM}
\bibfield{author}{\bibinfo{person}{Bank for International~Settlements}.} \bibinfo{year}{2014}\natexlab{}.
\newblock \bibinfo{title}{{Market-making and proprietary trading: industry trends, drivers and policy implications}}.
\newblock
\newblock
\showISBNx{ARRAY(0x63cd5310)}
\urldef\tempurl%
\url{https://ideas.repec.org/b/bis/biscgf/52.html}
\showURL{%
\tempurl}


\bibitem[\protect\citeauthoryear{Frank J.~Fabozzi}{Frank J.~Fabozzi}{2005}]%
        {Fabozzi}
\bibfield{author}{\bibinfo{person}{Steven~Mann Frank J.~Fabozzi}.} \bibinfo{year}{2005}\natexlab{}.
\newblock \bibinfo{booktitle}{\emph{The Handbook of Fixed Income Securities, Seventh Edition}}.
\newblock \bibinfo{publisher}{McGraw-Hill}, \bibinfo{address}{New York}.
\newblock


\bibitem[\protect\citeauthoryear{Fu, Peng, Ou, Sabharwal, and Khot}{Fu et~al\mbox{.}}{2023}]%
        {Fu2023}
\bibfield{author}{\bibinfo{person}{Yao Fu}, \bibinfo{person}{Hao Peng}, \bibinfo{person}{Litu Ou}, \bibinfo{person}{Ashish Sabharwal}, {and} \bibinfo{person}{Tushar Khot}.} \bibinfo{year}{2023}\natexlab{}.
\newblock \showarticletitle{{Specializing Smaller Language Models towards Multi-Step Reasoning}}.
\newblock \bibinfo{journal}{\emph{Proceedings of Machine Learning Research}}  \bibinfo{volume}{202} (\bibinfo{year}{2023}), \bibinfo{pages}{10421--10430}.
\newblock
\showISSN{26403498}
\showeprint[arxiv]{2301.12726}


\bibitem[\protect\citeauthoryear{Government}{Government}{2022}]%
        {AUDBudget}
\bibfield{author}{\bibinfo{person}{Australian Government}.} \bibinfo{year}{2022}\natexlab{}.
\newblock \bibinfo{title}{Australian Federal Budget 2022}.
\newblock
\newblock


\bibitem[\protect\citeauthoryear{Guo, Chen, Wang, Chang, Pei, Chawla, Wiest, and Zhang}{Guo et~al\mbox{.}}{2024}]%
        {GuoABMLLMijcai}
\bibfield{author}{\bibinfo{person}{Taicheng Guo}, \bibinfo{person}{Xiuying Chen}, \bibinfo{person}{Yaqi Wang}, \bibinfo{person}{Ruidi Chang}, \bibinfo{person}{Shichao Pei}, \bibinfo{person}{Nitesh~V. Chawla}, \bibinfo{person}{Olaf Wiest}, {and} \bibinfo{person}{Xiangliang Zhang}.} \bibinfo{year}{2024}\natexlab{}.
\newblock \showarticletitle{Large Language Model Based Multi-agents: A Survey of Progress and Challenges}. In \bibinfo{booktitle}{\emph{Proceedings of the Thirty-Third International Joint Conference on Artificial Intelligence, {IJCAI-24}}}, \bibfield{editor}{\bibinfo{person}{Kate Larson}} (Ed.). \bibinfo{publisher}{International Joint Conferences on Artificial Intelligence Organization}, \bibinfo{pages}{8048--8057}.
\newblock
\urldef\tempurl%
\url{https://doi.org/10.24963/ijcai.2024/890}
\showDOI{\tempurl}
\newblock
\shownote{Survey Track.}


\bibitem[\protect\citeauthoryear{Hayes, Todd, Chaidarun, Tepsuporn, Beling, and Scherer}{Hayes et~al\mbox{.}}{2014}]%
        {Hayes_2014}
\bibfield{author}{\bibinfo{person}{Roy Hayes}, \bibinfo{person}{Andrew Todd}, \bibinfo{person}{Nachapon Chaidarun}, \bibinfo{person}{Scott Tepsuporn}, \bibinfo{person}{Peter Beling}, {and} \bibinfo{person}{William Scherer}.} \bibinfo{year}{2014}\natexlab{}.
\newblock \bibinfo{title}{An agent-based financial simulation for use by researchers}.
\newblock
\newblock
\urldef\tempurl%
\url{https://doi.org/10.2139/ssrn.2423370}
\showDOI{\tempurl}


\bibitem[\protect\citeauthoryear{Hirchoua, Ouhbi, and Frikh}{Hirchoua et~al\mbox{.}}{2021}]%
        {HIRCHOUA2021114553}
\bibfield{author}{\bibinfo{person}{Badr Hirchoua}, \bibinfo{person}{Brahim Ouhbi}, {and} \bibinfo{person}{Bouchra Frikh}.} \bibinfo{year}{2021}\natexlab{}.
\newblock \showarticletitle{Deep reinforcement learning based trading agents: Risk curiosity driven learning for financial rules-based policy}.
\newblock \bibinfo{journal}{\emph{Expert Systems with Applications}}  \bibinfo{volume}{170} (\bibinfo{year}{2021}), \bibinfo{pages}{114553}.
\newblock
\showISSN{0957-4174}
\urldef\tempurl%
\url{https://doi.org/10.1016/j.eswa.2020.114553}
\showDOI{\tempurl}


\bibitem[\protect\citeauthoryear{Holland, Miller, and Miller}{Holland et~al\mbox{.}}{1991}]%
        {Holland}
\bibfield{author}{\bibinfo{person}{John~H. Holland}, \bibinfo{person}{John~H. Miller}, {and} \bibinfo{person}{John~H. Miller}.} \bibinfo{year}{1991}\natexlab{}.
\newblock \bibinfo{title}{Artificial Adaptive Agents in Economic Theory}.
\newblock
\newblock
\urldef\tempurl%
\url{https://doi.org/null}
\showDOI{\tempurl}


\bibitem[\protect\citeauthoryear{Ilut and Schneider}{Ilut and Schneider}{2022}]%
        {NBERw29915}
\bibfield{author}{\bibinfo{person}{Cosmin~L Ilut} {and} \bibinfo{person}{Martin Schneider}.} \bibinfo{year}{2022}\natexlab{}.
\newblock \bibinfo{booktitle}{\emph{Modeling Uncertainty as Ambiguity: a Review}}.
\newblock \bibinfo{type}{Working Paper} 29915. \bibinfo{institution}{National Bureau of Economic Research}.
\newblock
\urldef\tempurl%
\url{https://doi.org/10.3386/w29915}
\showDOI{\tempurl}


\bibitem[\protect\citeauthoryear{Imani and Du}{Imani and Du}{2023}]%
        {Imani2023}
\bibfield{author}{\bibinfo{person}{Shima Imani} {and} \bibinfo{person}{Liang Du}.} \bibinfo{year}{2023}\natexlab{}.
\newblock \showarticletitle{{MathPrompter: Mathematical Reasoning using Large Language Models}}.
\newblock \bibinfo{journal}{\emph{Proceedings of the Annual Meeting of the Association for Computational Linguistics}} \bibinfo{volume}{5}, \bibinfo{number}{1} (\bibinfo{year}{2023}), \bibinfo{pages}{37--42}.
\newblock
\showISBNx{9781959429685}
\showISSN{0736587X}
\urldef\tempurl%
\url{https://doi.org/10.18653/v1/2023.acl-industry.4}
\showDOI{\tempurl}
\showeprint[arxiv]{2303.05398}


\bibitem[\protect\citeauthoryear{Kirilenko, Kyle, Samadi, and Tuzun}{Kirilenko et~al\mbox{.}}{2018}]%
        {Kirilenko2018}
\bibfield{author}{\bibinfo{person}{Andrei~A. Kirilenko}, \bibinfo{person}{Albert~S. Kyle}, \bibinfo{person}{Mehrdad Samadi}, {and} \bibinfo{person}{Tugkan Tuzun}.} \bibinfo{year}{2018}\natexlab{}.
\newblock \showarticletitle{{Automation, Intermediation and the Flash Crash}}.
\newblock \bibinfo{journal}{\emph{null}} \bibinfo{volume}{16}, \bibinfo{number}{4} (\bibinfo{year}{2018}), \bibinfo{pages}{17--28}.
\newblock
\urldef\tempurl%
\url{https://doi.org/10.2139/ssrn.3119363}
\showDOI{\tempurl}


\bibitem[\protect\citeauthoryear{Lewkowycz, Andreassen, Dohan, Dyer, Michalewski, Ramasesh, Slone, Anil, Schlag, Gutman-Solo, et~al\mbox{.}}{Lewkowycz et~al\mbox{.}}{2022}]%
        {lewkowycz2022solving}
\bibfield{author}{\bibinfo{person}{Aitor Lewkowycz}, \bibinfo{person}{Anders Andreassen}, \bibinfo{person}{David Dohan}, \bibinfo{person}{Ethan Dyer}, \bibinfo{person}{Henryk Michalewski}, \bibinfo{person}{Vinay Ramasesh}, \bibinfo{person}{Ambrose Slone}, \bibinfo{person}{Cem Anil}, \bibinfo{person}{Imanol Schlag}, \bibinfo{person}{Theo Gutman-Solo}, {et~al\mbox{.}}} \bibinfo{year}{2022}\natexlab{}.
\newblock \showarticletitle{Solving quantitative reasoning problems with language models}.
\newblock \bibinfo{journal}{\emph{Advances in Neural Information Processing Systems}}  \bibinfo{volume}{35} (\bibinfo{year}{2022}), \bibinfo{pages}{3843--3857}.
\newblock


\bibitem[\protect\citeauthoryear{Luo, Sun, Xu, Zhao, Lou, Tao, Geng, Lin, Chen, and Zhang}{Luo et~al\mbox{.}}{2023}]%
        {Luo2023}
\bibfield{author}{\bibinfo{person}{Haipeng Luo}, \bibinfo{person}{Qingfeng Sun}, \bibinfo{person}{Can Xu}, \bibinfo{person}{Pu Zhao}, \bibinfo{person}{Jianguang Lou}, \bibinfo{person}{Chongyang Tao}, \bibinfo{person}{Xiubo Geng}, \bibinfo{person}{Qingwei Lin}, \bibinfo{person}{Shifeng Chen}, {and} \bibinfo{person}{Dongmei Zhang}.} \bibinfo{year}{2023}\natexlab{}.
\newblock \bibinfo{title}{{WizardMath: Empowering Mathematical Reasoning for Large Language Models via Reinforced Evol-Instruct}}.
\newblock
\newblock
\showeprint[arxiv]{2308.09583}
\urldef\tempurl%
\url{http://arxiv.org/abs/2308.09583}
\showURL{%
\tempurl}


\bibitem[\protect\citeauthoryear{Massa and Simonov}{Massa and Simonov}{2003}]%
        {MASSA200399}
\bibfield{author}{\bibinfo{person}{Massimo Massa} {and} \bibinfo{person}{Andrei Simonov}.} \bibinfo{year}{2003}\natexlab{}.
\newblock \showarticletitle{Reputation and interdealer trading: a microstructure analysis of the Treasury Bond market}.
\newblock \bibinfo{journal}{\emph{Journal of Financial Markets}} \bibinfo{volume}{6}, \bibinfo{number}{2} (\bibinfo{year}{2003}), \bibinfo{pages}{99--141}.
\newblock
\showISSN{1386-4181}
\urldef\tempurl%
\url{https://doi.org/10.1016/S1386-4181(02)00045-9}
\showDOI{\tempurl}


\bibitem[\protect\citeauthoryear{Mehrab, Stundal, Swarup, Venaktramanan, Lewis, Mortveit, Barrett, Pandey, Wells, Galvani, Singer, Leblang, Colwell, and Marathe}{Mehrab et~al\mbox{.}}{2024}]%
        {SamarthNetwork}
\bibfield{author}{\bibinfo{person}{Zakaria Mehrab}, \bibinfo{person}{Logan Stundal}, \bibinfo{person}{Samarth Swarup}, \bibinfo{person}{Srinivasan Venaktramanan}, \bibinfo{person}{Bryan Lewis}, \bibinfo{person}{Henning Mortveit}, \bibinfo{person}{Christopher Barrett}, \bibinfo{person}{Abhishek Pandey}, \bibinfo{person}{Chad Wells}, \bibinfo{person}{Alison Galvani}, \bibinfo{person}{Burton Singer}, \bibinfo{person}{David Leblang}, \bibinfo{person}{Rita Colwell}, {and} \bibinfo{person}{Madhav Marathe}.} \bibinfo{year}{2024}\natexlab{}.
\newblock \showarticletitle{Network Agency: An Agent-based Model of Forced Migration from Ukraine}. In \bibinfo{booktitle}{\emph{Proceedings of the 23rd International Conference on Autonomous Agents and Multiagent Systems}} (Auckland, New Zealand) \emph{(\bibinfo{series}{AAMAS '24})}. \bibinfo{publisher}{International Foundation for Autonomous Agents and Multiagent Systems}, \bibinfo{address}{Richland, SC}, \bibinfo{pages}{1372–1380}.
\newblock
\showISBNx{9798400704864}


\bibitem[\protect\citeauthoryear{of~Australia}{of~Australia}{2021}]%
        {RBA}
\bibfield{author}{\bibinfo{person}{Reserve~Bank of Australia}.} \bibinfo{year}{2021}\natexlab{}.
\newblock \bibinfo{title}{Government-bond-markets-in-advanced-economies-during-the-pandemic}.
\newblock , \bibinfo{numpages}{52 - 60}~pages.
\newblock
\urldef\tempurl%
\url{www.rba.gov.au/publications/bulletin/2021/sep/government-bond-markets-in-advanced-economies-during-the-pandemic.html}
\showURL{%
\tempurl}


\bibitem[\protect\citeauthoryear{of~England}{of~England}{2015}]%
        {BOE}
\bibfield{author}{\bibinfo{person}{Bank of England}.} \bibinfo{year}{2015}\natexlab{}.
\newblock \bibinfo{title}{GEMM Guidebook: A guide to the roles of the DMO and Primary Dealers in the UK Government bond market}.
\newblock
\newblock
\urldef\tempurl%
\url{https://dmo.gov.uk/media/ownmub2i/guidebook20150401.pdf}
\showURL{%
\tempurl}


\bibitem[\protect\citeauthoryear{of~International~Settlements}{of~International~Settlements}{2023}]%
        {BIS2023}
\bibfield{author}{\bibinfo{person}{Bank of International~Settlements}.} \bibinfo{year}{2023}\natexlab{}.
\newblock \bibinfo{title}{{BIS Quarterly Review, December 2023}}.
\newblock
\newblock
\showISBNx{9789292597139}


\bibitem[\protect\citeauthoryear{Park, O'Brien, Cai, Morris, Liang, and Bernstein}{Park et~al\mbox{.}}{2023}]%
        {park2023generative}
\bibfield{author}{\bibinfo{person}{Joon Park}, \bibinfo{person}{Joseph O'Brien}, \bibinfo{person}{Carrie Cai}, \bibinfo{person}{Meredith~Ringel Morris}, \bibinfo{person}{Percy Liang}, {and} \bibinfo{person}{Michael Bernstein}.} \bibinfo{year}{2023}\natexlab{}.
\newblock \bibinfo{title}{Generative Agents: Interactive Simulacra of Human Behavior}.
\newblock
\newblock
\showeprint[arxiv:2304.03442]{2304.03442}~[cs.HC]
\urldef\tempurl%
\url{https://arxiv.org/abs/2304.03442}
\showURL{%
\tempurl}


\bibitem[\protect\citeauthoryear{Paulin, Calinescu, and Wooldridge}{Paulin et~al\mbox{.}}{2019}]%
        {Paulin_2019}
\bibfield{author}{\bibinfo{person}{James Paulin}, \bibinfo{person}{Anisoara Calinescu}, {and} \bibinfo{person}{Michael Wooldridge}.} \bibinfo{year}{2019}\natexlab{}.
\newblock \showarticletitle{Understanding flash crash contagion and systemic risk: A micro–macro agent-based approach}.
\newblock \bibinfo{journal}{\emph{Journal of Economic Dynamics and Control}}  \bibinfo{volume}{100} (\bibinfo{date}{March} \bibinfo{year}{2019}), \bibinfo{pages}{200–229}.
\newblock
\showISSN{0165-1889}
\urldef\tempurl%
\url{https://doi.org/10.1016/j.jedc.2018.12.008}
\showDOI{\tempurl}


\bibitem[\protect\citeauthoryear{Pinter}{Pinter}{2023}]%
        {Pinter2023}
\bibfield{author}{\bibinfo{person}{Gabor Pinter}.} \bibinfo{year}{2023}\natexlab{}.
\newblock \bibinfo{title}{{An anatomy of the 2022 gilt market crisis}}.
\newblock
\newblock
\urldef\tempurl%
\url{https://doi.org/10.2139/ssrn.4488189}
\showDOI{\tempurl}


\bibitem[\protect\citeauthoryear{Pinter, Wang, and Zou}{Pinter et~al\mbox{.}}{2024}]%
        {Pinter2024}
\bibfield{author}{\bibinfo{person}{Gabor Pinter}, \bibinfo{person}{Chaojun Wang}, {and} \bibinfo{person}{Junyuan Zou}.} \bibinfo{year}{2024}\natexlab{}.
\newblock \showarticletitle{{Size Discount and Size Penalty: Trading Costs in Bond Markets}}.
\newblock \bibinfo{journal}{\emph{Review of Financial Studies}} \bibinfo{volume}{37}, \bibinfo{number}{7} (\bibinfo{year}{2024}), \bibinfo{pages}{2156--2190}.
\newblock
\showISSN{14657368}
\urldef\tempurl%
\url{https://doi.org/10.1093/rfs/hhae007}
\showDOI{\tempurl}


\bibitem[\protect\citeauthoryear{Rudiger, Rudiger, Vigier, and Vigier}{Rudiger et~al\mbox{.}}{2019}]%
        {Rudiger_2019}
\bibfield{author}{\bibinfo{person}{Jesper Rudiger}, \bibinfo{person}{Jesper Rudiger}, \bibinfo{person}{Adrien Vigier}, {and} \bibinfo{person}{Adrien Vigier}.} \bibinfo{year}{2019}\natexlab{}.
\newblock \bibinfo{title}{Who Acquires Information in Dealer Markets}.
\newblock
\newblock
\urldef\tempurl%
\url{https://doi.org/10.2139/ssrn.2875763}
\showDOI{\tempurl}


\bibitem[\protect\citeauthoryear{Securities and Commission}{Securities and Commission}{2013}]%
        {SEC_13TradeOrderVolumeRatios}
\bibfield{author}{\bibinfo{person}{U.S. Securities} {and} \bibinfo{person}{Exchange Commission}.} \bibinfo{year}{2013}\natexlab{}.
\newblock \bibinfo{title}{Trade and Order Volume Ratios}.
\newblock
\newblock
\urldef\tempurl%
\url{https://www.sec.gov/about/trade-order-volume-ratios}
\showURL{%
\tempurl}
\newblock
\shownote{Accessed: 2024-09-10.}


\bibitem[\protect\citeauthoryear{Srivastava, Rastogi, Rao, Shoeb, Abid, and et~al}{Srivastava et~al\mbox{.}}{2023}]%
        {srivastava2023imitationgamequantifyingextrapolating}
\bibfield{author}{\bibinfo{person}{Aarohi Srivastava}, \bibinfo{person}{Abhinav Rastogi}, \bibinfo{person}{Abhishek Rao}, \bibinfo{person}{Abu Awal~Md Shoeb}, \bibinfo{person}{Abubakar Abid}, {and} \bibinfo{person}{Adam~Fisch et al}.} \bibinfo{year}{2023}\natexlab{}.
\newblock \bibinfo{title}{Beyond the Imitation Game: Quantifying and extrapolating the capabilities of language models}.
\newblock
\newblock
\showeprint[arxiv]{2206.04615}~[cs.CL]
\urldef\tempurl%
\url{https://arxiv.org/abs/2206.04615}
\showURL{%
\tempurl}


\bibitem[\protect\citeauthoryear{{U.S. Securities and Exchange Commission}}{{U.S. Securities and Exchange Commission}}{2024}]%
        {SEC_24ExchangeTradeOrderVolume}
\bibfield{author}{\bibinfo{person}{{U.S. Securities and Exchange Commission}}.} \bibinfo{year}{2024}\natexlab{}.
\newblock \bibinfo{title}{Market Structure: Exchange Trade Volume}.
\newblock
\newblock
\urldef\tempurl%
\url{https://www.sec.gov/marketstructure/datavis}
\showURL{%
\tempurl}
\newblock
\shownote{Accessed: 2024-09-10.}


\bibitem[\protect\citeauthoryear{Vadori, Ardon, Ganesh, Spooner, Amrouni, Vann, Xu, Zheng, Balch, and Veloso}{Vadori et~al\mbox{.}}{2022}]%
        {vadori2022multiagent}
\bibfield{author}{\bibinfo{person}{Nelson Vadori}, \bibinfo{person}{Leo Ardon}, \bibinfo{person}{Sumitra Ganesh}, \bibinfo{person}{Thomas Spooner}, \bibinfo{person}{Selim Amrouni}, \bibinfo{person}{Jared Vann}, \bibinfo{person}{Mengda Xu}, \bibinfo{person}{Zeyu Zheng}, \bibinfo{person}{Tucker Balch}, {and} \bibinfo{person}{Manuela Veloso}.} \bibinfo{year}{2022}\natexlab{}.
\newblock \bibinfo{title}{Towards Multi-Agent Reinforcement Learning driven Over-The-Counter Market Simulations}.
\newblock
\newblock
\showeprint[arxiv]{2210.07184}~[cs.MA]


\bibitem[\protect\citeauthoryear{Varghese and Edward}{Varghese and Edward}{2018}]%
        {doi:10.1177/2277975217733858}
\bibfield{author}{\bibinfo{person}{Jose Varghese} {and} \bibinfo{person}{Manoj Edward}.} \bibinfo{year}{2018}\natexlab{}.
\newblock \showarticletitle{Relationship Between Job Orientation and Performance of Sales People: A Financial Services Industry Perspective}.
\newblock \bibinfo{journal}{\emph{IIM Kozhikode Society \& Management Review}} \bibinfo{volume}{7}, \bibinfo{number}{1} (\bibinfo{year}{2018}), \bibinfo{pages}{88--96}.
\newblock
\urldef\tempurl%
\url{https://doi.org/10.1177/2277975217733858}
\showDOI{\tempurl}


\bibitem[\protect\citeauthoryear{Vidler and Walsh}{Vidler and Walsh}{2024a}]%
        {vidler2024tradertalkllmbehaviouralabm}
\bibfield{author}{\bibinfo{person}{Alicia Vidler} {and} \bibinfo{person}{Toby Walsh}.} \bibinfo{year}{2024}\natexlab{a}.
\newblock \showarticletitle{{ TraderTalk: An LLM Behavioural ABM applied to Simulating Human Bilateral Trading Interactions }}. In \bibinfo{booktitle}{\emph{2024 IEEE International Conference on Agents (ICA)}}. \bibinfo{publisher}{IEEE Computer Society}, \bibinfo{address}{Los Alamitos, CA, USA}, \bibinfo{pages}{164--167}.
\newblock
\urldef\tempurl%
\url{https://doi.org/10.1109/ICA63002.2024.00042}
\showDOI{\tempurl}


\bibitem[\protect\citeauthoryear{Vidler and Walsh}{Vidler and Walsh}{2024b}]%
        {vidler2024decodingotcgovernmentbond}
\bibfield{author}{\bibinfo{person}{Alicia Vidler} {and} \bibinfo{person}{Toby Walsh}.} \bibinfo{year}{2024}\natexlab{b}.
\newblock \bibinfo{title}{Decoding OTC Government Bond Market Liquidity: An ABM Model for Market Dynamics}.
\newblock
\newblock
\showeprint[arxiv]{2501.16331}~[q-fin.TR]
\urldef\tempurl%
\url{https://arxiv.org/abs/2501.16331}
\showURL{%
\tempurl}


\bibitem[\protect\citeauthoryear{Vidler and Walsh}{Vidler and Walsh}{2025}]%
        {vidler2025evaluatingbinarydecisionbiases}
\bibfield{author}{\bibinfo{person}{Alicia Vidler} {and} \bibinfo{person}{Toby Walsh}.} \bibinfo{year}{2025}\natexlab{}.
\newblock \bibinfo{title}{Evaluating Binary Decision Biases in Large Language Models: Implications for Fair Agent-Based Financial Simulations}.
\newblock
\newblock
\showeprint[arxiv]{2501.16356}~[cs.LG]
\urldef\tempurl%
\url{https://arxiv.org/abs/2501.16356}
\showURL{%
\tempurl}


\bibitem[\protect\citeauthoryear{Wei, Tay, Bommasani, Raffel, Zoph, Borgeaud, Yogatama, Bosma, Zhou, Metzler, Chi, Hashimoto, Vinyals, Liang, Dean, and Fedus}{Wei et~al\mbox{.}}{2022a}]%
        {Wei2022a}
\bibfield{author}{\bibinfo{person}{Jason Wei}, \bibinfo{person}{Yi Tay}, \bibinfo{person}{Rishi Bommasani}, \bibinfo{person}{Colin Raffel}, \bibinfo{person}{Barret Zoph}, \bibinfo{person}{Sebastian Borgeaud}, \bibinfo{person}{Dani Yogatama}, \bibinfo{person}{Maarten Bosma}, \bibinfo{person}{Denny Zhou}, \bibinfo{person}{Donald Metzler}, \bibinfo{person}{Ed~H. Chi}, \bibinfo{person}{Tatsunori Hashimoto}, \bibinfo{person}{Oriol Vinyals}, \bibinfo{person}{Percy Liang}, \bibinfo{person}{Jeff Dean}, {and} \bibinfo{person}{William Fedus}.} \bibinfo{year}{2022}\natexlab{a}.
\newblock \bibinfo{title}{Emergent Abilities of Large Language Models}.
\newblock
\newblock
\showeprint[arxiv]{2206.07682}~[cs.CL]
\urldef\tempurl%
\url{https://arxiv.org/abs/2206.07682}
\showURL{%
\tempurl}


\bibitem[\protect\citeauthoryear{Wei, Wang, Schuurmans, Bosma, Ichter, Xia, Chi, Le, and Zhou}{Wei et~al\mbox{.}}{2022b}]%
        {ChainOfThoughtWei2022}
\bibfield{author}{\bibinfo{person}{Jason Wei}, \bibinfo{person}{Xuezhi Wang}, \bibinfo{person}{Dale Schuurmans}, \bibinfo{person}{Maarten Bosma}, \bibinfo{person}{Brian Ichter}, \bibinfo{person}{Fei Xia}, \bibinfo{person}{Ed~H. Chi}, \bibinfo{person}{Quoc~V. Le}, {and} \bibinfo{person}{Denny Zhou}.} \bibinfo{year}{2022}\natexlab{b}.
\newblock \showarticletitle{{Chain-of-Thought Prompting Elicits Reasoning in Large Language Models}}.
\newblock \bibinfo{journal}{\emph{Advances in Neural Information Processing Systems}} \bibinfo{volume}{35}, \bibinfo{number}{NeurIPS} (\bibinfo{year}{2022}), \bibinfo{pages}{1--43}.
\newblock
\showISBNx{9781713871088}
\showISSN{10495258}
\showeprint[arxiv]{2201.11903}


\bibitem[\protect\citeauthoryear{Xu, Tao, Shen, Xu, Xu, Long, and guang Lou}{Xu et~al\mbox{.}}{2024}]%
        {xu2024rereadingimprovesreasoninglarge}
\bibfield{author}{\bibinfo{person}{Xiaohan Xu}, \bibinfo{person}{Chongyang Tao}, \bibinfo{person}{Tao Shen}, \bibinfo{person}{Can Xu}, \bibinfo{person}{Hongbo Xu}, \bibinfo{person}{Guodong Long}, {and} \bibinfo{person}{Jian guang Lou}.} \bibinfo{year}{2024}\natexlab{}.
\newblock \bibinfo{title}{Re-Reading Improves Reasoning in Large Language Models}.
\newblock
\newblock
\showeprint[arxiv]{2309.06275}~[cs.CL]
\urldef\tempurl%
\url{https://arxiv.org/abs/2309.06275}
\showURL{%
\tempurl}


\bibitem[\protect\citeauthoryear{Zamfirescu-Pereira, Wong, Hartmann, and Yang}{Zamfirescu-Pereira et~al\mbox{.}}{2023}]%
        {Zamfirescu-Pereira2023}
\bibfield{author}{\bibinfo{person}{J.D. Zamfirescu-Pereira}, \bibinfo{person}{Richmond~Y. Wong}, \bibinfo{person}{Bjoern Hartmann}, {and} \bibinfo{person}{Qian Yang}.} \bibinfo{year}{2023}\natexlab{}.
\newblock \showarticletitle{Why Johnny Can’t Prompt: How Non-AI Experts Try (and Fail) to Design LLM Prompts}. In \bibinfo{booktitle}{\emph{Proceedings of the 2023 CHI Conference on Human Factors in Computing Systems}} (Hamburg, Germany) \emph{(\bibinfo{series}{CHI '23})}. \bibinfo{publisher}{Association for Computing Machinery}, \bibinfo{address}{New York, NY, USA}, Article \bibinfo{articleno}{437}, \bibinfo{numpages}{21}~pages.
\newblock
\showISBNx{9781450394215}
\urldef\tempurl%
\url{https://doi.org/10.1145/3544548.3581388}
\showDOI{\tempurl}


\bibitem[\protect\citeauthoryear{Zhang, Bosselut, Yasunaga, Ren, Liang, Manning, and Leskovec}{Zhang et~al\mbox{.}}{2022}]%
        {zhang2022}
\bibfield{author}{\bibinfo{person}{Xikun Zhang}, \bibinfo{person}{Antoine Bosselut}, \bibinfo{person}{Michihiro Yasunaga}, \bibinfo{person}{Hongyu Ren}, \bibinfo{person}{Percy Liang}, \bibinfo{person}{Christopher~D. Manning}, {and} \bibinfo{person}{Jure Leskovec}.} \bibinfo{year}{2022}\natexlab{}.
\newblock \bibinfo{title}{{Greaselm: Graph Reasoning Enhanced Language Models for Question Answering}}.
\newblock , \bibinfo{numpages}{16}~pages.
\newblock
\showeprint[arxiv]{2201.08860}


\bibitem[\protect\citeauthoryear{Zhang, Mao, Ge, Wang, Wynter, Xia, and Wu}{Zhang et~al\mbox{.}}{2024}]%
        {zhang2024llmmastermindsurveystrategic}
\bibfield{author}{\bibinfo{person}{Yadong Zhang}, \bibinfo{person}{Shaoguang Mao}, \bibinfo{person}{Tao Ge}, \bibinfo{person}{Xun Wang}, \bibinfo{person}{Adrian Wynter}, \bibinfo{person}{Yan Xia}, {and} \bibinfo{person}{Wenshan Wu}.} \bibinfo{year}{2024}\natexlab{}.
\newblock \bibinfo{title}{LLM as a Mastermind: A Survey of Strategic Reasoning with Large Language Models}.
\newblock
\newblock
\showeprint[arxiv]{2404.01230}~[cs.CL]
\urldef\tempurl%
\url{https://arxiv.org/abs/2404.01230}
\showURL{%
\tempurl}


\end{thebibliography}

\end{document}